\documentclass[preprint2]{aastex}

\begin{document}

\title{Rapid rotation of ultra-Li-depleted halo stars and their association
with blue stragglers
\footnote{Based on observations obtained with the University College London 
\'echelle spectrograph (UCLES) on the Anglo-Australian Telescope (AAT) and with the
Utrecht \'echelle spectrograph (UES) on the William Herschel Telescope (WHT).}
\\
{\normalsize To appear in The Astrophysical Journal, tentatively v. 571 (20 May 2002)}
}

\author{
Sean G. Ryan\altaffilmark{1},
Scott G. Gregory\altaffilmark{1},
Ulrich Kolb  \altaffilmark{1},
Timothy C. Beers\altaffilmark{2},
Toshitaka Kajino\altaffilmark{3}
}
\altaffiltext{1}{Department of Physics and Astronomy, The Open University, Walton Hall,
Milton Keynes, MK7 6AA, UK; email: s.g.ryan@open.ac.uk; s.gregory@open.ac.uk; u.c.kolb@open.ac.uk}
\altaffiltext{2}{Department of Physics and Astronomy, Michigan State University, 
East Lansing, MI 48824-1116; email: beers@pa.msu.edu }
\altaffiltext{3}{National Astronomical Observatory, Mitaka, Tokyo, 181-8588 Japan,
and University of Tokyo; 
email: kajino@ferio.mtk.nao.ac.jp}

\begin{abstract} 
Observations of eighteen halo main-sequence-turnoff stars, four of which are extremely
deficient in Li, show that three of the Li-poor ones
have substantial line broadening. We attribute this to stellar rotation. Despite the
great ages of halo stars, for G202-65, BD+51$^\circ$1817 and Wolf~550 we infer 
$v$sin$i$ = 8.3$\pm$0.4,
7.6$\pm$0.3,
and 5.5$\pm$0.6~km~s$^{-1}$ respectively. The stated errors are 3$\sigma$.
For CD$-$31$^\circ$19466 we derive a 3$\sigma$ upper limit $v$sin$i$ $<$ 2.2~km~s$^{-1}$.
The three rotating stars are known spectroscopic binaries. We explain the high rotation
velocities in terms of mass and angular momentum transfer onto the surface
of the turnoff star from an initially more-massive donor. 
Estimates of the specific angular momentum of accreted material indicate
that quite small transfer masses could have been involved, though the 
unknown subsequent spin-down of the accretor prevents us from
assigning definitive values for each star.
The accretor is now seen
as an ultra-Li-deficient star whose origin makes it a low-mass counterpart of field 
blue stragglers. The Li could have been destroyed before or during the mass-transfer
episode. Such objects must be avoided in studies of the primordial Li abundance and
in investigations into the way normal single stars process their initial Li.

\end{abstract}

\ \\
{Suggested short title:}  \\
Rotation of ultra-Li-depleted stars 


\section{Introduction}\label{sec:intro}

The lithium content of halo stars near the main-sequence turnoff is of great
importance for several reasons. The near-constancy of the Li abundances, 
which are broadly independent of 
metallicity or effective temperature, led \citet{ss82} to conclude that these were
hardly altered from the primordial value. That discovery prompted numerous studies
over the ensuing two decades, with the aim of using the inferred primordial abundance
as a constraint on the baryon density of the Universe, $\Omega_{\rm B}$
(e.g. \citet{rbofn00}).

Lithium is also important because it is a sensitive probe of mixing below the
stellar surface. Since Li is destroyed in stars 
at the relatively low temperature of 
2.5$\times 10^6$~K, it survives in halo main-sequence turnoff stars
only in a thin surface layer making up a few percent of the stellar mass. 
Its destruction
in cooler dwarfs (e.g. \citet{rd98}, \citet{clm99}), 
and subgiants \citep{psb93, rd95}
provides constraints on stellar structure and evolution models that cannot at 
present be obtained by any other means. 
By combining observations with stellar evolutionary models, \citet{pwsn01} 
infer that even main-sequence turnoff Pop~II stars have destroyed 
$\simeq 0.2$~dex
of their Li, though \citet{rnb99} infer only $<$0.1~dex.
The difference in these values
depends on (A) the frequency of the few stars with Li abundances significantly
lower than the rest, (B) the distribution of their levels of Li deficiency, and
(C) whether the deviations are in fact due to the proposed mechanism
(rotationally-induced mixing) and not some other.
These constraints directly impact on the inferred
primordial abundance.
Small amounts of Li {\it production} may also be present in the stars
that formed more recently from nucleosynthetically enriched material. This
can be tracked analogously to the way He production is tracked by d$Y$/d$Z$.

In stars where some Li is clearly post-primordial, such as where 
$^6$Li indicates Galactic production, or where 
elevated $^7$Li abundances argue strongly for 
post-big bang production, Li provides important constraints on a range of 
nucleosynthesis mechanisms and their associated sites. These include cosmic ray 
spallation (and fusion) reactions, production in supernovae via the 
$\nu$-process,
and production in stellar sources such as asymptotic giant branch (AGB) stars,
novae, and 
red-giants (e.g. \citet{rmmb99},  \citet{rkbsrmr01} and references therein).

Although the majority of halo main-sequence turnoff stars have almost identical
Li abundances, about 7\% have very low (thus far undetected) Li abundances at 
least a factor of five below the plateau values expected for their 
temperatures and metallicities \citep{sms84, hm91, hwt91, t92, smfs93, rbkr01}. 
Likewise, a similar fraction of disk stars appears to be
anomalously Li-deficient \citep{lhe91, rbkr01}. It has been unclear why these
small numbers of stars should differ so significantly from the Li-normal stars. 
Their evolutionary states \citep{t94}, and the presence (or lack) of abundance 
anomalies for other elements \citep{nrbd97, rnb98} have thus far failed to
provide unambiguous evidence of their origin.
Citing the common features of Li deficiency and a clustering towards the 
main-sequence turnoff,  \citet{rbkr01} suggested a common origin with 
field blue stragglers. They proposed that the same mechanism may produce both
classes of objects, and that the different names which have been assigned
historically --- blue stragglers and ultra-Li-deficient stars--- depend only 
on whether they have super-turnoff or sub-turnoff masses.
In this and what follows, a distinction is made between
cluster blue stragglers and their field counterparts, because while the former are
believed to arise predominantly from collisions, possibly leading to stellar 
mergers, the latter are more likely the result of
mass transfer \citep{ffpb95, mhno90, ps00}.

\citet{rkbsrmr01} analysed the Li abundances of 18 halo main-sequence turnoff 
stars, and found that four of them are
ultra-Li-deficient objects. During detailed spectral analysis of other elements
\citep{rgkb02} we recognised that three of the Li-depleted stars,
but none of
the Li-normal stars, exhibit unusually broad absorption lines. We
believe this is due to rotational broadening. In this paper we present 
measurements of the projected rotation velocities and discuss the implications 
of this discovery for the origin of these systems and their possible link
with field blue stragglers.

\section{Observations and Measurements}\label{sec:obs}

\subsection{Spectra}

The observational programme has been described already by \citet{rkbsrmr01},
so only 
key details will be repeated here. 
We had noted previously that studies of the possible dependence of the halo
Li abundance on metallicity and effective temperature (a surrogate for mass)
were potentially weakened by a non-uniform sampling of these parameters.
We set out to improve the situation by observing halo 
stars with higher than average metallicity and effective temperature, since these
were the types of stars that had been overlooked.
The program stars were selected on the basis of 
effective temperatures and metallicities estimated from photometric colours and in some 
cases 1~\AA-resolution blue spectra. Nothing was known about their Li abundances.
They were expected to have 6000~K~$^<_\sim$~$T_{\rm eff}$~$^<_\sim$~6400~K and 
$-2~ ^<_\sim$~[Fe/H]~$^<_\sim -1$, and hence provide a supplemental sample of 
metal-deficient turnoff stars that probed a region of metallicity and
temperature space that had been poorly investigated.

Spectra with 
typical S/N = 50-90 per 0.05~\AA\ pixel
were obtained with the University College London 
\'echelle spectrograph (UCLES) on the Anglo-Australian Telescope 
at a resolving power 
$\lambda/\Delta \lambda = 43000$ ($FWHM$ = 7.0~km~s$^{-1}$)
and with the
Utrecht \'echelle spectrograph (UES) on the William Herschel Telescope
at a resolving power 
$\lambda/\Delta \lambda = 50000$ ($FWHM$ = 6.0~km~s$^{-1}$).
The quoted velocity widths are from measurements of ThAr lines. 
Importantly, we used
observations of ThAr lamps to confirm 
that the spectrograph resolution was stable through the night.

The spectra
included the Li~{\small I} 6707~\AA\ line, and extended to shorter wavelengths
where lines of heavier metals could be observed.
Equivalent widths $W_{\rm G}$ were measured
by fitting Gaussian profiles using IRAF's splot ``kk'' facility, 
and are tabulated by \citet{rgkb02}. In the same operation, the 
fitted Gaussian's $FWHM$ was also recorded.

\subsection{Line profiles}

The measurements of Fe~I lines were scrutinised for potential blends by 
examining the dependence of $FWHM$ on $W_{\rm G}$.
The expectation was that lines formed from the 
blending of two or more components would have unusually large Gaussian widths
for their equivalent widths. However, comparison of the data for the various
stars showed that the Li-deficient objects had systematically broader
line profiles than the normal stars.

\begin{figure}
\plotone{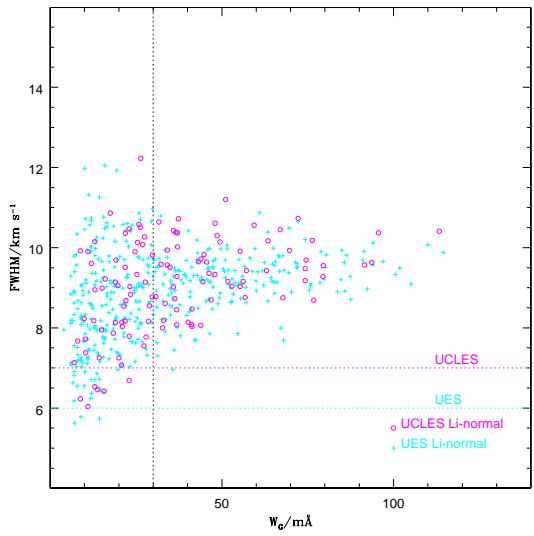}
\caption{Gaussian fit parameters for Fe~I lines in Li-normal stars.
{\it pluses}: UES data for Li-normal stars.
{\it circles}: UCLES data for Li-normal stars.
The vertical dashed line at 30~m\AA\ indicates a low-$W_{\rm G}$ cutoff,
below
which noise dominates and we neglect the measurements.
The horizontal dashed lines give the instrumental FWHM of the two spectrograph
configurations.
}
\label{FigA}
\end{figure}

Figure~\ref{FigA} 
shows $FWHM$ of the Fe~I lines of 
Li-normal stars as a function
of the measured equivalent width. Lines with equivalent widths in the range
30~$^<_\sim$~$(W_{\rm G}/{\rm m\AA})$~$^<_\sim$~120 exhibit a slight increase 
of $FWHM$
from
9 to 10~km~s$^{-1}$ as the spectral lines strengthen and their Gaussian cores 
saturate. 
The intrinsic $FWHM$ of the lines due to natural, thermal,
microturbulent, and collisional broadening is only 3-4~km~s$^{-1}$,
so the observed value is dominated by the finite spectral resolution of the
adopted spectrograph setup (6.0 and 7.0 km~s$^{-1}$), plus a contribution
from macroturbulence which we discuss below.
Note that the UCLES and UES data are just distinguishable, consistent with the
slightly better spectral resolution employed with UES.
Lines weaker than 30~m\AA\ are considerably affected by noise, such that the
diagram fans out to higher and lower $FWHM$. 

For the weakest lines, 
$FWHM$  should 
depend only weakly on $W_{\rm G}$, but it is likely that a
selection bias de-populates the top-left of the
diagram. Lines whose errors conspire to produce shallow lines would give rise to
small equivalent widths and large $FWHM$, but these are more likely to
be overlooked in the measurement process.
Lines with the same equivalent width but smaller $FWHM$ will be deeper
and more likely to be measured.
There is some evidence for such a bias in the figure. Because of this and 
the increasing relative errors in the weakest lines, we
do not rely on the profile measurements of lines weaker
than 30~m\AA\ in the subsequent analysis.

Figure~\ref{FigE} 
extends Figure~\ref{FigA} with the inclusion of the four ultra-Li-poor stars.
CD$-$31$^\circ$19466 
is indistinguishable from Li-normal stars similarly observed with UCLES,
but the other three, all observed with UES, have broader lines than 
Li-normal UES stars.
The figure also
shows linear least-squares fits to the subset of lines with equivalent widths 
$W_{\rm G} > 30$~m\AA. The fits are listed in Table~\ref{TabA}. 
Clearly an additional broadening mechanism
is at work in three of the stars. The fact that three out of four Li-poor stars show
this broadening, whereas zero out the sixteen Li-normal stars show the effect, strongly
suggests that the broadening is related to the extreme Li deficiency of these objects.

\section{Analysis}

\subsection{Line broadening mechanisms}

In this subsection we consider mechanisms that might be capable of producing the 
excess broadening in the Li-poor stars.
If it was produced by a Gaussian mechanism,
then the additional broadening required, $FWHM$$_{\rm +}$,
could be estimated as the difference (in quadrature)
between the Gaussian $FWHM$ measured for the Li-depleted stars and 
that for the Li-normal stars. 
By evaluating the difference in the least-squares fits 
at $W_{\rm G} = 50$~m\AA\ we obtain the broadening velocities
listed in Table~\ref{TabA}.

One Gaussian broadening mechanism is microturbulence, but we note at the outset 
that extreme values of microturbulence are not responsible. Quite apart from the 
uncommonly large velocities that
would be required, the abundance analysis \citep{rgkb02} resulted in normal
microturbulent velocities being derived for these stars, $\xi \simeq 1$~km~s$^{-1}$.

An alternative mechanism is macroturbulence. The radial-tangential assumption 
for macroturbulence produces a non-Gaussian broadening function having
a sharp core and non-zero wings that are less pronounced than those of a 
Gaussian (e.g. \citet{g92}).
Measurements of macroturbulence have rarely been made in halo stars, 
since the values are so low 
that very high resolving powers are essential. However, they have
been measured in analyses of the $^6$Li/$^7$Li isotope ratio in a few stars.
In their analysis of HD~76932,  \citet{nls94} used a
1D model atmosphere to derive a 
radial-tangential macroturbulent component with $FWHM$~=~4.7~km~s$^{-1}$.
For a larger sample of halo turnoff stars, \citet{sln98} derived Gaussian
macroturbulent values,
$FWHM$$_{\Gamma}$,
in the range 
4.4-6.1~km~s$^{-1}$.
In their analysis of G271-162, \citet{nahd00} derived
two measures separately for 1D and 3D model atmospheres.
The former required $FWHM$$_{\Gamma}$ = 5.9$\pm$0.1~km~s$^{-1}$.
In the 3D hydrodynamical model, the microturbulent and macroturbulent velocity
fields are inherent in the model, and only a small rotational component 
($v$sin$i$ = 2.9$\pm$0.1~km~s$^{-1}$) needed to be added.

The observed $FWHM$ values for the Li normal stars could be explained
by macroturbulence of the magnitude found by \citet{sln98}.
For example, if lines with $W_{\rm G}$~=~50~m\AA\ have
intrinsic widths with $FWHM_{\rm int.}$~=~3.5~km~s$^{-1}$, 
experience Gaussian macroturbulence with 
$FWHM_{\Gamma}$~=~6.0~km~s$^{-1}$,
and encounter Gaussian instrumental profiles with
$FWHM_{\rm inst.}$~=~6.0 and 7.0~km~s$^{-1}$,
then we would expect to observe $FWHM$~=~9.2 and 9.9~km~s$^{-1}$
respectively.
This is consistent with the values observed for the Li-normal stars.
In contrast, very large $FWHM$$_{\Gamma}$
values would be required to match the Li-poor ones. 
Li-poor stars with 
intrinsic widths of $FWHM$~=~3.5~km~s$^{-1}$,
and observed on UES at an instrumental $FWHM_{\rm inst.}$~=~6.0~km~s$^{-1}$
would require 
$FWHM_{\Gamma}$ = 8.5~km~s$^{-1}$ to reach an observed value
$FWHM$ = 11.0~km~s$^{-1}$, and would require
$FWHM_{\Gamma}$ = 11.6~km~s$^{-1}$ to reach an observed value
$FWHM$ = 13.5~km~s$^{-1}$.
Although these values are large compared with ones found previously,
we should bear in mind that
Li-depleted stars are {\it not} normal, so we should remain open to the
possibility that they may have unusually turbulent atmospheres. 
It would be valuable to obtain higher S/N spectra to test whether the line 
profiles
could be fit by macroturbulence. If they could,
the turbulence may be related to a mixing process which led to the
destruction of the surface Li. 
That is, if large scale motions in the
stellar atmosphere have carried surface Li to depths great enough to destroy it,
then perhaps we are seeing that motion directly. It would remain, however, 
to identify the mechanism responsible for
driving that motion, and how it might be sustained to the present in halo stars
with ages $\simeq$12-14 Gyr.

Another alternative, which we consider more likely,
is that the broadening is due to rotation.
Normal halo dwarfs do not rotate perceptibly.
There are few measurements of rotation in halo main-sequence
stars, again because it is generally below the level of detectability 
at the resolving power $R < 40000$ at which these stars are typically observed.
It is for this reason
that we are forced to use the fourteen Li-normal stars in our sample as the
non-rotating reference stars.
However, \citet{sln98} do provide guideline measurements for nine halo stars,
for which they find $v$sin$i$~$\le$~2--3~km~s$^{-1}$.
The reason for the low rotation of halo dwarfs is
that they have presumably spun
down by magnetic breaking over their long lifetimes. However, 
the Li-depleted stars clearly are peculiar, and it is quite possible
that three of them have high rotation velocities.

A rotational broadening
profile is non-Gaussian, so the rotation velocity cannot be derived as easily as
the Gaussian broadening velocity $FWHM$$_{\rm +}$ above,
but nevertheless, it must be of the 
same order of magnitude. Rotation estimates are made below. 
We defer our discussion of the implications of measured rotation velocities to 
Section~4,
and instead continue here by explaining how velocity measurements were made
from the broadened spectral lines.

\subsection{Rotational broadening measurements}

\citet{g92} derives an analytic broadening function for the case where rotational
broadening dominates a line profile. Gray suggests a value $v$sin$i$ $>$ 15~km~s$^{-1}$
for this condition to be satisfied. Although for our stars we infer slower rotation
than this, $\simeq$~5.5-8.3~km~s$^{-1}$, spectrum synthesis suggests that the 
non-rotating intrinsic line width, $FWHM_{\rm int.}$,
is only 3-4~km~s$^{-1}$, 
so rotation apparently 
does dominate. It is possible that our derived rotation velocities are
slightly in error because of this simplification,  
but the resulting (small) discrepancy should not affect our conclusions.

The broadening profile due to stellar rotation has two components, one
elliptical and one parabolic.
In contrast to Gaussian, Lorentzian, Voigt and macroturbulent profiles, the rotational
profile has
no wings beyond $\Delta \lambda$ = $\pm$ $\lambda v$sin$i$/$c$, 
so the resulting profile is a sharp U-shape, scaled in wavelength 
by the projected equatorial rotation velocity $v$sin$i$.
The relative contribution of the two parts of the profile
depends on the limb darkening of the star, which enters Gray's formula via a parameter
$\eta$, which we set equal to 0.6. The model assumes no differential rotation of the
stellar surface layers.

\begin{figure}
\plotone{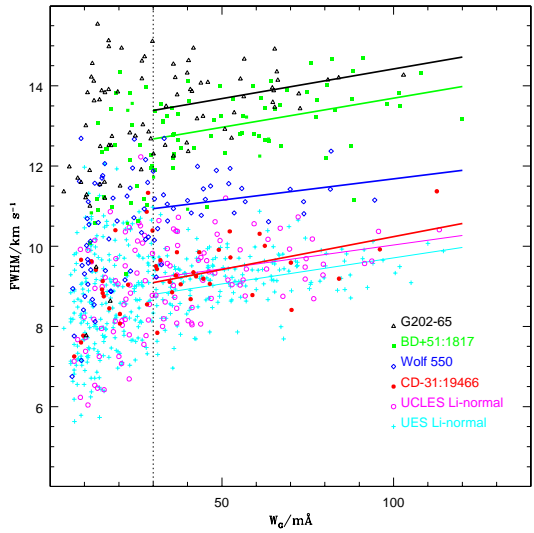}
\caption{Gaussian fit parameters for Fe~I lines in all stars.
{\it pluses}: UES data for Li-normal stars.
{\it open circles}: UCLES data for Li-normal stars.
{\it other symbols}: Li-poor stars.
{\it heavy solid lines}: Least-squares fits to lines with $W > 30$~m\AA\ in
(from top to bottom) G202-65, BD+51$^\circ$1817, Wolf~550, and 
CD$-$31$^\circ$19466.
{\it light solid lines}: Least-squares fits to lines with 
$W_{\rm G} > 30$~m\AA\ in
Li-normal stars observed with UCLES (upper) and UES (lower).
}
\label{FigE}
\end{figure}

The rotation velocities of the Li-deficient stars were calculated as follows.
Synthetic spectra were computed using code that originated with 
\citet{cn78}.
The stellar atmosphere parameters and models were the same
as adopted by \citet{rkbsrmr01}, and the line list was the same set of \ion{Fe}{1}
lines measured by \citet{rgkb02} and used in Figure~\ref{FigE}. 

The synthetic spectra were first broadened by Gaussian profiles intended to match the
instrumental and macroturbulent broadening, 
and then the spectral lines were measured using the same IRAF
Gaussian-fitting routine as was used for the observed spectra. In this way, 
$FWHM$ and $W_{\rm G}$ values were obtained for the synthetic spectra that 
could
be compared with those from the observations.

The $FWHM$ of the synthesised lines is not uniquely a 
function of $W_{\rm G}$, but also depends slightly on the lower excitation 
potential $\chi_{\rm low}$ of the transition.
This dependence reflects the greater depths at which lines
of higher excitation potential form (e.g. Gray 1992, his Fig. 13.2 and 13.3).
Because of this depth dependence, two lines having the same equivalent width 
$W$ but different excitation potentials will have different profiles and 
hence different values of $W_{\rm G}$ 
(which is the Gaussian approximation to the equivalent width $W$). 
It is for these reasons that it is important to use the same spectral lines
in the synthetic spectra as in the observations.\footnote{This dependence of
the line profile on $\chi_{\rm low}$ as well as $W$ is of course important
to consider when constraining the instrumental and macroturbulent broadening 
of other line profiles, for example of Li and Ba when attempting to constrain 
their isotopic composition.}
Note that the range in  $FWHM$ brought about by the $\chi_{\rm low}$
dependence is less than the observed spread, which is to say that random errors
in line measurements dominate the observed spread and could be reduced with 
higher S/N data.

The Gaussian broadening was
varied until the run of $FWHM$ vs $W_{\rm G}$ 
matched that seen in Figure~\ref{FigE} for the Li-normal stars.
Specifically, a least-squares fit was performed on the measurements of the synthesised
lines, just as had been done for the real data, and the two least-squares fits were
compared at $W = 50$~m\AA. This choice of line strength was made because such lines
are
strong enough to avoid much of the noise that affects weaker lines in the observed
spectra, and are central enough in the distributions of both $W_{\rm G}$ and
$FWHM$ to avoid being grossly affected by outlying data.

The Gaussian FWHM required to achieve this match was
8.1$\pm$0.1~km~s$^{-1}$ for the UCLES spectrum and 
7.8$\pm$0.1~km~s$^{-1}$ for the UES spectrum.
Subtracting (in quadrature) the known underlying instrumental values implies 
average macroturbulent values $FWHM_\Gamma$~=~4.1~km~s$^{-1}$ for the UCLES
sample and 5.0~km~s$^{-1}$ for the UES sample.
The stated 1$\sigma$ uncertainties reflect the standard error in the mean 
of the observed $FWHM$ values, based on the scatter about the 
least-squares fit.
These instrumental and macroturbulent components
supplement the natural, thermal, collisional, and microturbulent broadening
incorporated in the synthesis. 

\begin{figure}
\plotone{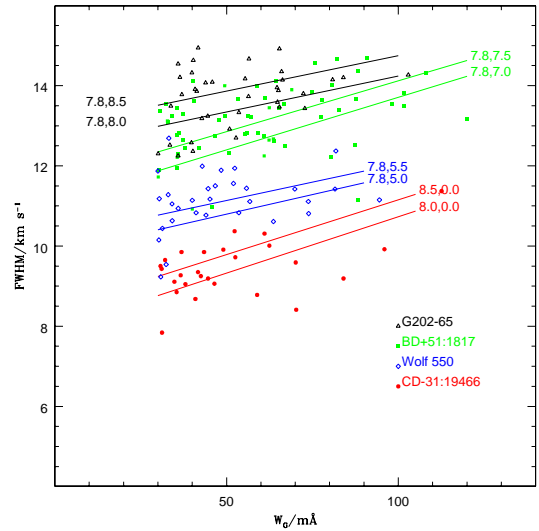}
\caption{Least-squares fits to Gaussian parameters of
synthetic spectra, for the stated broadening velocities (in km~s$^{-1}$).
The first velocity is the FWHM of instrumental and macroturbulent origin, 
and the second is $v$sin$i$.
The observed data are also shown.
}
\label{FigD}
\end{figure}

Once the line broadening of the Li-normal stars was reproduced, 
the Gaussian components were frozen
and the rotational broadening was increased from zero to determine what value 
of $v$sin$i$ was required to reproduce the run of $FWHM$ vs $W_{\rm G}$ 
in the Li-depleted stars.  The set of lines synthesised was tuned
from star to star to match the list of lines actually observed. The
synthesised lines were then measured using the 
same Gaussian fitting routine as for the observations, and a least-squares fit was
made to the run of $FWHM$ vs $W_{\rm G}$ for the lines with 
$W_{\rm G} > 30$~m\AA.
Figure~\ref{FigD} shows the least-squares fits for the synthetic spectra,
along with the observational data for the Li-poor stars.
The match between the observed and synthesized fits was again
made at $W_{\rm G} = 50$~m\AA.
The derived $v$sin$i$  velocities are given in column~(6) of Table~\ref{TabA}.
The tabulated uncertainty is the change in $v$sin$i$ required to alter
$FWHM$ (at $W_{\rm G} = 50$~m\AA) by three times the value $RMS/\sqrt{n}$,
where $n$ is the number of spectral lines in the least-squares
fit and $RMS$ is the scatter about the fit.
Similarly, the upper limit on $v$sin$i$ for CD$-31^\circ19466$ is the
change in $v$sin$i$ required to alter
$FWHM$ (at $W_{\rm G} = 50$~m\AA) by three times the value $RMS/\sqrt{n}$.

We note for completeness that the run of $FWHM$ vs $W_{\rm G}$
for the synthetic spectra was slightly steeper than for the observed
lines. The origin of this mismatch could be a combination of the
effects of errors near the 30~m\AA\ cutoff and imperfections in the
line profiles used in the synthetic spectra, such as deficiencies in the various
broadening functions used.
This is unlikely to have had a major impact on the derived velocities
because the action of matching the observed and theoretical
least-squares fits at a ``central''
value of the range of parameters (i.e. 50~m\AA)
will minimise the impact of errors in the slope.

\begin{figure}
\plotone{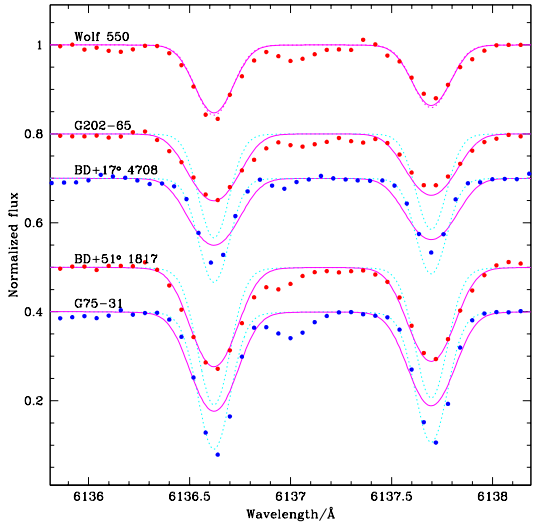}
\caption{Observed and synthetic spectra of two \ion{Fe}{1} lines for the 
Li-poor stars Wolf~550, G202-65 and BD+51$^\circ$1817. 
All synthetic spectra have been broadened by a Gaussian with
$FWHM$~=~7.8~km~s$^{-1}$ to reproduce the instrumental and macroturbulent
broadening.
For Wolf~550, rotational broadening with $v$sin$i$ = 5.0 and 5.5~km~s$^{-1}$ 
has been included. The two synthetic profiles are difficult
to distinguish, but nevertheless have significantly different $FWHM$
values --- see Figure~\ref{FigD}. 
The synthetic spectra for G202-65 have $v$sin$i$ = 0 and 8.5~km~s$^{-1}$, 
and for BD+51$^\circ$1817 have $v$sin$i$ = 0 and 7.5~km~s$^{-1}$. 
These pairs of syntheses are also shown superimposed
on the spectra of the Li-normal stars BD+17$^\circ$4708 and G75-31 
respectively.
}
\label{FigF}
\end{figure}

As a final illustration of the broadening, we show in Figure~\ref{FigF}
a sample of spectra around two \ion{Fe}{1} lines. 
The spectrum of Wolf~550 is overlaid by two synthetic spectra having 
$v$sin$i$ = 5.0 and 5.5~km~s$^{-1}$. The difference between the two line
profiles is very difficult to discern on any single absorption line, but
Figure~\ref{FigD} indicates that the $FWHM$ values are nevertheless
very sensitive to this small change. It is because we have used data for a 
large number of \ion{Fe}{1} lines that we have been able to achieve 
this level of sensitivity. Moreover, even though rotation velocities
this small are challenging to measure, the key implications of this
work follow from the fact that they are rotating significantly at all, 
rather than from the precise value of the velocity that should be assigned.

The spectrum of G202-65 in Figure~\ref{FigF} is also overlaid by two
synthetic spectra, the deeper one having a Gaussian broadening of 
$FWHM$~=~7.8~km~s$^{-1}$ but no
rotation broadening, and the shallow one having the same Gaussian broadening
and $v$sin$i$~=~8.5~km~s$^{-1}$.
These same synthetic spectra are also overlaid on the spectrum of the
Li-normal star BD+17$^\circ$4708 which has similar line strengths to
G202-65, but has no discernible rotation broadening. The differences in the
non-rotating and rotating profiles are clear.

A similar comparison is made in the case of BD+51$^\circ$1817, where
the deeper spectrum again has Gaussian broadening of 
$FWHM$~=~7.8~km~s$^{-1}$ but no
rotation broadening, and the shallow profile has the same Gaussian broadening
and $v$sin$i$~=~7.5~km~s$^{-1}$.
These synthetic spectra are overlaid on the spectrum of the
Li-normal star G75-31, which has similar line strengths to
BD+51$^\circ$1817 but no discernible rotation broadening.
The comparison makes clear that the profiles give reasonable fits to the
spectra, but we emphasise again that the greatest sensitivity comes from 
measuring the FWHM of a larger number of lines, as shown in Figures~\ref{FigE}
and \ref{FigD}.

We note that independently, \citet{cllgm01} have assigned 
``marginally significant detections''
of rotation velocities for these stars between 6 and 10~km~s$^{-1}$.
Their data, taken with S/N from 7-50 per resolution element 
(5-35 per pixel?), did not allow them to achieve the same precision as
here, but the similarity of their ``marginal'' detections to the more
precise results derived here verify their ability to identify the presence of
rotational broadening and derive approximate
rotation velocities even from low S/N spectra.

\subsection{Metallicities}

The metallicities of the stars estimated from photometry and
low-resolution spectroscopy \citep{rkbsrmr01}
have been improved using the high-resolution spectra. That analysis is discussed
more fully by \citet{rgkb02}. When the original photometric estimates of 
$T_{\rm eff}$ and surface gravities log~$g$~=~4.2 were adopted, the calculated
metallicities were in good agreement with those originally inferred. 
For the 18 stars in the original programme, the mean difference was
$\langle$[Fe/H]$_{\rm new}-$[Fe/H]$_{\rm old}\rangle$~=~$0.02$~dex, 
with a standard deviation 0.14~dex,
consistent with the error estimates presented originally.
Both the old and new values are listed for the Li-poor stars
in Table~\ref{TabA}, columns (8) and (9).

\subsection{Effective temperatures}

Calculations were also made of the effective temperatures that would be
inferred for each star by nulling the dependence of derived abundance 
on excitation potential. For the 18 stars, $T_{\rm eff}$ increased by
on average 246~K from this process, though with a broad standard deviation
of 179~K. The inferred metallicities increased accordingly.
However, for two of the hot Li-depleted stars the inferred $T_{\rm eff}$
increased by more than 500~K, implying $T_{\rm eff} \sim 6900$~K.
Such high temperatures would push those stars clear of the main sequence
turnoff, and would greatly strengthen their association with traditional
blue stragglers. 
If their
temperatures are that high, they would almost certainly be genuine blue 
stragglers and not correctly be associated with sub-turnoff-mass objects.
However, even in that case we would have two new problems to consider:
why the photometric colours of these stars would be incompatible with
their effective temperatures, and how many other super-turnoff-mass 
blue stragglers might also be masquerading as sub-turnoff-colour stars.
We note that the photometric effective temperatures derived
by us \citep{rkbsrmr01} are consistent with those of \citet{cllgm01},
though both are based on similar photometry.

Alternative diagnostics of the effective temperature may help
resolve this problem.
Two of the Li-poor stars have available 1--2 \AA\  resolution spectroscopy
obtained during the course of the HK survey of Beers and colleagues 
(see \citet{brnrs99} and references therein).  
A calibration of $T_{\rm eff}$ 
based on the strength of a Balmer-line index (HP2), and tied to the
infra-red flux method (IRFM) scale of  
\citet{aamr96},
yields temperature estimates
$T_{\rm eff} = 6223 \pm 125$ K for BD+51$^\circ$1817, and 
$T_{\rm eff} = 6286 \pm 125$ K for Wolf 550.  
Furthermore, Wolf 550 (= G66-30) is one of the stars measured by
\citet{aamr96},
who report $T_{eff} = 6211 \pm 107$ K.  All of
these results are consistent with the temperatures we derived from 
photometric measurements, thus strengthening our belief that they, rather than
the spectroscopic determinations, are appropriate.
Nevertheless, the higher effective temperatures and corresponding 
metallicities that would be inferred from the spectral analysis are also 
given in the table (columns (10) and (11)).

\section{Discussion}

\subsection{Characterising the stars}

Besides rotation, there is one additional difference, and possibly two, between
the three broadened Li-depleted stars and the one non-broadened one. First, the
three broadened stars are very close to the halo main-sequence turnoff, having 
$6269$~K~$\le$~$T_{\rm eff}$(photometric)~$\le$~6390~K, whereas 
CD$-$31$^\circ$19466 is cooler, with $T_{\rm eff}$ = 5986~K \citep{rbkr01}. 
Second, 
\citet{clla94}, \citet{cllgm01}
and Latham (2000, private communication) list
all three of the hot group as confirmed binaries.
Their periods $P_{\rm orb}$ (where only the last digit is uncertain) and 
eccentricities $e$ are listed in Table~\ref{TabA}.
The binary status of the non-rotating Li-depleted star is not yet known.

As the $v$sin$i$ measurements are lower limits on the equatorial
rotation velocities $v$,
we can derive upper limits on the rotation periods. The sidereal period of the
Sun is 25.4~days, giving $v_\odot$~=~2.0~km~s$^{-1}$. The projected
surface velocities of the rotating Li-poor stars are typically three times 
higher, and their zero-age main-sequence (ZAMS) radii 
can be estimated for inferred masses of 0.8~M$_\odot$ as
$R_{\rm ZAMS} \simeq 0.8$R$_\odot$, assuming Pop~I structures. 
Halo stars, having lower opacity, will be slightly more 
compact than Pop~I stars for a given mass. 
On the other hand, stars at the
turnoff will have larger radii than ZAMS objects,
being more evolved.
The ZAMS radius
would lead to rotation periods of typically only 7~days; 
even a factor-of-two increase in radius
would increase the rotation periods to only 14~days. 
These rotation periods are well below the current
orbital periods, and hence the orbits are not currently tidally synchronised.

\subsection{Evolutionary histories and spin-up}

If rotation
is responsible for the broadening, then one of two histories may be invoked:
(a) these stars formed with unusually high rotation velocities $\simeq$14~Gyr ago and
have not yet spun down, or
(b) they have acquired angular momentum during their lifetimes.

Option (a) is rather ad hoc; while the details of the way stars spin down is 
not understood completely (e.g. \citet{spt00}),
it is unclear how a small fraction of the halo population
could avoid the spin down that every other halo dwarf seems to have experienced.

Option (b), on the other hand, could result from a range of mechanisms which are
already well studied, such as mass transfer onto the star from
a companion, an event which also transfers angular momentum, or via spin up in a 
merger. These mechanisms have already been proposed by
\citet{rbkr01} as possibly responsible for the formation of the ultra-Li-deficient
stars; that work preceded the detection of rotation.

The three rotating stars we have investigated are similar in many respects to 
the thick-disk, binary blue straggler HR~4657 identified by \citet{fb99}.
\citet{fb99} likewise drew a connection between rapid rotation,
Li depletion, and mass transfer from a companion.
Our present results strengthen the evidence for this connection,
which we now make in three additional cases,
and which we explore in more detail below.
We note that our sample also extends this connection from the thick disk to 
the halo, since the space velocities of G202-65 and Wolf~550 are indisputably
those of halo stars, even if BD+51$^\circ$1817 might be more correctly 
associated with the thick disk (\citet{rkbsrmr01}, their Table 2).

The possibility that  the broadened stars acquired additional
angular momentum via mass transfer seems particularly plausible given the
confirmed binary status of all three of these objects, since we can then
point to a likely
mass donor. Moreover, \citet{cllgm01} infer projected companion masses 
$M_2$sin$i$~$<$~0.55~$M_\odot$,
compatible with their being the white-dwarf remnants of stars initially 
more massive than the 0.8~$M_\odot$ normally associated with the halo 
main-sequence turnoff.

The current rotation velocities of the three stars are well below the high 
values that are {\it initially} obtained during mergers. 
Calculations by \citet{w76}
showed 
velocities $>100$~km~s$^{-1}$ maintained throughout the main-sequence 
lifetime of a merger product, dropping only to
$v_{\rm rot}$ = 89.4~km~s$^{-1}$ by the time it reached the base of the giant 
branch.  More recent work 
\citep{slbdrs97, ll95}
raised the possibility that such objects would spin down
during hydrogen burning if the merger product goes through a period
of expansion and subsequent re-contraction that resembles pre-main-sequence
evolution, which is when most of the spin-down of single low-mass stars occurs
\citep{spt00}.
The low velocities offer no support for an unrelaxed merger of the type studied
by Webbink, but cannot rule out a merger which subsequently spins down
as modelled by Leonard \& Livio.
However, the need for a merger history is obviated by the existence of 
a present companion, which is an alternative angular-momentum donor. 

\citet{hept95} examine four binary evolutionary channels by which
matter could be  
transferred onto stars. Investigating Ba and CH stars, they find that
mass transfer via Roche-lobe overflow 
leads to $10^2$ to $10^3$~day periods. 
(They studied both stable and unstable transfer, the latter involving 
common-envelope ejection.)
On the other hand, accretion from a 
stellar wind (``wind accretion'' and ``wind exposure'')
leads to periods $10^3$ to $10^5$~days.
Stars that experience Roche-lobe overflow are expected to circularize their orbits
on a short timescale. However, \citet{cllgm01} find significantly non-zero 
eccentricities for two of the three binary systems,
Wolf 550 and G202-65, which makes it unlikely that
Roche-lobe overflow occurred in these systems. If the initially more-massive 
companion 
was always well inside its Roche lobe, tidal circularisation would have been
ineffective 
(e.g.\ \citet{z77}),
and the observed eccentricities would
effectively be primordial. 

An alternative possibility is that the orbits {\it were} circularized in the 
past,
but the sudden mass loss from the system in 
a subsequent supernova explosion caused non-zero 
present-day eccentricities.  
In this case,
the invisible companion would be a neutron star, and
its likely mass would be $1.4 M_\sun$,
since most measured neutron-star masses sit very close to this value
(e.g.\ \citet{c98}). 
However, the measured mass functions would then imply uncomfortably-small 
inclination angles for at
least two of the three systems (see column (15) of Table~\ref{TabA}). 
As any Type~II
supernova explosion is likely to have 
ejected more than twice the current binary mass ($\simeq 2 M_\sun$),
the system could not have survived a spherically-symmetric
explosion where the post--supernova orbital speed is larger than
the escape speed. However, survival would be possible if the neutron star 
received a
suitably-directed kick velocity at birth. Such kicks are
inferred from the observed proper motions of pulsars, e.g.\ 
\citep{k96}.

A phase of Roche-lobe overflow may have occurred in 
BD+51$^\circ$1817, whose eccentricity is much lower.
The present system parameters are consistent with the remnant
configuration of a stable mass-transfer phase with 
a low--mass giant donor. In that case the companion would be a white
dwarf, and the orbital period should be consistent with an approximate
orbital-period--white-dwarf-mass relation 
(e.g.\ \citet{wrs83})
that reflects the 
radius ($R_2$) -- core mass ($M_c$) relation of low--mass giant stars. 
For moderate Population II metallicity $Z=0.001$ this is  
\begin{equation}
\frac{R_2}{R_\sun} = 10^{3.36} \times \left( \frac{M_c}{M_\sun} \right)^{4.17}
\end{equation}
(e.g.\ \citet{r99}).
Combining this with Kepler's third law for a semi-detached system  
\begin{equation}
\log_{10} {{P_{\rm orb}}\over{\rm days}} = 1.5 \times \log_{10} \left( \frac{R_2}{R_\sun} \right) -
           0.5 \times \log_{10} \left( \frac{M_2}{M_\sun} \right) - 0.433
\end{equation}
we can express the orbital period of a Roche lobe-filling giant
in terms of its mass and core mass. Assuming mass transfer removes the entire
envelope (so that $M_2 = M_c$) gives
the white-dwarf-remnant mass 
$M_2 \simeq 0.47 M_\sun$, consistent  with the mass function measured
by \citet{cllgm01}. This mass would imply an orbital inclination
$i \simeq 48^\circ$.  

The progenitor of 
BD+51$^\circ$1817
could have escaped unstable mass transfer
only if the initial mass ratio was not far from unity, so that 
the initially more massive star can shed enough mass in a wind on
the AGB to become the less massive component at Roche-lobe
contact. This is required because convective stars such as giants expand
upon mass loss, and only if 
the mass ratio $q = M_{\rm donor}/M_{\rm accretor}$ is sufficiently
small can the Roche lobe expand fast enough to keep the
expanding star in check. For conservative mass transfer from a 
fully-convective star this stability limit is $q \la 2/3$, 
e.g.\ \citet{w85}.

We can estimate an upper limit on the transferred mass
$\Delta M$ in the phase of stable Roche-lobe overflow by assuming that
the evolution was 
conservative, i.e.\ proceeded with constant binary mass and constant
orbital angular momentum. In that case 
the product $M_1^3 M_2^3 P_{\rm orb}$ is constant throughout the
evolution ($M_1$ is the accretor mass). 
If the mass ratio {\it at the beginning of Roche-lobe overflow} was close to
the critical value 2/3, the current masses ($M_1=0.90 M_\sun$, $M_2=0.47
M_\sun$) then imply initial masses $M_{1,i}=0.82 M_\sun$, $M_{2,i}=0.55
M_\sun$, i.e.\ $\Delta M \la 0.08 M_\sun$. 
(Of course, the donor would have
been even more massive during its main-sequence lifetime,
before non-conservative mass loss via winds set in.)

If on the other hand the mass ratio at Roche-lobe contact was larger
than the critical value, the progenitor of BD+51$^\circ$1817 would have
experienced dynamically-unstable mass transfer from
a giant star, leading to a short--lived common-envelope phase. 
The orbit tightens as orbital energy is used to eject the envelope of
the giant star 
(e.g.\ \citet{il93} for a review). 
In this scenario,
the fact that
the post--common-envelope orbit is still fairly wide implies 
that the binding energy of the giant immediately before Roche-lobe 
contact was small. 

Given these possibilities for the evolutionary history, the three
Li--depleted halo stars could have accreted mass and thus
acquired angular momentum. The circular orbit of BD+51$^\circ$1817 allows this
to have occurred during Roche-lobe overflow or 
a common-envelope phase, whereas the other stars with
non-zero eccentricities are more compatible with accretion from
a stellar wind.

Although it is clear that the stars are not tidally synchronised in their 
{\it present} orbits, we briefly considered the possibility that the rapid 
spin of these binary components
could be left over from earlier tidal synchronisation 
when the orbital separation was much less. This would require that the stars
once had orbital periods as short as 7-14~days, were synchronized at that time,
and subsequently migrated apart.
Clearly this is not possible for BD+51$^\circ$1817 if the companion is
a white dwarf. If there was stable Roche-lobe overflow the initial
orbital period must have been longer than 400 days; if there was a
common enevelope phase the initial period must have been even longer
than the current period. 
Wind losses could have increased the orbital period, but not by such a large
factor.
If the angular momentum per unit mass in the wind is the specific
orbital angular momentum of the corresponding star, the orbital
period increases as $P_{\rm orb} \propto 1/M^2$ (where $M$ is the total binary
mass). Hence,
to increase the binary period from $14$~days to the
current values, the total binary mass must have decreased by a factor
between 3  and 7, i.e.\ by more than $\simeq 5 M_\sun$. Stars are thought
to
lose that much mass only in the late stages of their evolution when
they are large --- too large to fit inside the small orbit that
corresponds to $P_{\rm orb} = 14$~days.
If, on the other hand, the present companion of either of the three
systems considered here is a neutron star, then the orbital period could
have been as short as 14 days after a common-envelope phase that
removed the hydrogen 
envelope of the progenitor of the neutron star. However, such a tight
system would then not be able to increase its orbital separation
through the supernova, as even in the presence of kick velocities the 
orbital separation cannot increase by more than a factor 2
(\citet{k96}).  
We conclude that none of the three systems discussed here ever had an
orbital period as short as 14 days.

\subsection{Mass and angular momentum estimates}

\subsubsection{Roche-lobe overflow}

In the case of Roche-lobe overflow, an accretion disc would form since
the circularisation radius
is typically a tenth of the orbital separation and hence larger than the
likely stellar radius of the accretor in most cases. If the 
Keplerian disc 
extends down to the surface of the accretor, the disc material 
carries the angular momentum per unit mass, $j$, given by
\begin{equation}
j = \sqrt{GM_1R_1}
\label{jkepler}
\end{equation}
where $M_1$ and $R_1$ is the mass and radius of the accretor. 
Then the gain of angular momentum, $\Delta J$, is
\begin{equation}
\Delta J = \int \dot M_1 j {\rm d}t = \int j {\rm d}M_1 .
\end{equation}
Assuming a simple main--sequence mass--radius relation 
\begin{equation}
\frac{R_1}{R_\sun} = r_0 \, \frac{M_1}{M_\sun}
\end{equation}
(where $r_0 \simeq 1$) this becomes
\begin{equation}
\Delta J = \frac{1}{2} \sqrt{\frac{GR_1}{M_1}} \left(M_1^2 - M_{1,i}^2 \right), 
\end{equation}
where $M_{1,i}$ is the initial accretor mass. Writing the accretor's
moment of inertia as $k^2 M_1 R_1^2$ (where $k^2 \simeq 0.1$) and assuming
that the initial spin is zero gives
\begin{equation}
\frac{M_{1,i}}{M_1} = \left( 1 - 2 k^2 \frac{P_k}{P} \right)^{1/2},
\label{spinup}
\end{equation}
where $P$ is the current spin period, and $P_k$ the ``break--up'' spin
period   
\begin{equation}
P_k = \frac{2\pi}{\sqrt{GM_1/R_1^3}} \simeq 0.1 {\rm days}.
\end{equation}
Expression~\ref{spinup} shows that the accretion of only $\Delta M = M_1
- M_{1,i} \simeq 10^{-3} M_\sun$  spins up the accretor to rates 
comparable to the currently observed spin period $P \simeq 7$~days.   
If the rapid rotation is restricted to the convective envelope of the
accretor, the required accreted mass is only $\simeq 10^{-4}
M_\sun$
(with $k^2 \simeq 0.01$ in the expression for the moment of inertia,
from e.g.\ our own Population II stellar models, calculated with
Mazzitelli's stellar code; \citet{m89}).

As we measure only the projected velocity $v$sin$i$, the actual velocities
may be greater; the periods may then
be shorter than $P \simeq 7$~days. If the period were only 
$P \simeq 1-2$ days, as would also be the case if the companion is
a neutron star at low inclination (see Table~\ref{TabA}), we would infer 
$\Delta M = 10^{-2} M_\sun$.  

It is clear from these analyses that accretion of quite small masses would
be sufficient to spin up these objects.

\subsubsection{Slow wind accretion}

In the case of accretion from a sufficiently slow wind of a
companion that is not too far from its Roche lobe an accretion disc
may still form, so that the same estimate for $\Delta J$ and $\Delta
M$ applies. 

\subsubsection{Fast wind accretion}

In the case of a fast wind the accretion flow can be
described as Bondi--Hoyle accretion. Here the specific angular
momentum $j$ of the accreted material is difficult to estimate;
3-D hydrodynamical simulations (Ruffert 1999) show that $j$ is a
fraction $f \simeq 0 - 0.7$ of the total angular momentum per unit mass 
available
in the accretion cylinder. If the accretion radius is of order the
stellar radius this translates into a relation similar to
(\ref{spinup}), but with a coefficient $\simeq k^2/f$ instead of
$k^2$. So, for example, if $f=0.1$ the accreted mass would have to 
be $\Delta M \simeq 10^{-2} M_\sun$. 
Given the small capture efficiency and short lifetime of phases with
large wind loss rates the total wind--accreted mass is
typically $\la 0.1 M_\sun$, which is adequate, but could be much 
smaller 
(e.g.\ \citet{kk99}).

\subsubsection{Common-envelope phase}

In the case of a common-envelope phase the net mass gain
is probably small, perhaps $\la 10^{-2} M_\sun - 0.1 M_\sun$ 
\citep{ht91}. 
The specific angular momentum of the accreted
material is presumably similar to the Bondi--Hoyle case.

\subsubsection{Caveats}

The calculations above assume that the stars have not spun down
significantly following the mass-transfer event.
If the currently observed spin is due to an episode of mass
accretion several Gyrs ago it is likely that the stars have spun down
significantly via magnetic stellar-wind braking. Hence our estimates
for the accreted mass must be seen as lower limits.

\subsection{Li depletion in a mass-transfer binary}

The link between Li depletion and rapid rotation is easy to envisage in a mass-transfer
scenario. Revisiting discussions by \citet{nrbd97} and \citet{rbkr01}, at least 
three 
possibilities exist if the rapid rotation reflects angular-momentum transfer onto the
turnoff star via mass accretion.
(1) Triggered mixing: depending on how the surface of the
accretor is disturbed by the accreting plasma, additional mixing may be induced that
carries the surface Li to depths where it is destroyed.
(2) Prior destruction: even if no substantial mixing was triggered in the 
accretion event, both stars may have been devoid of Li already once mass 
transfer began. In this scenario, the donor would destroy its original Li during
its giant branch evolution (or earlier), and/or the accretor may have initially
been of sufficiently low mass ($M$~$^<_\sim$~0.7~$M_\odot$) that it too had 
already destroyed its Li by conventional means.
(3) Accretion-dominated surface: if the mass accreted is higher than the
the mass of the accretor's surface convection zone, then the composition of the
material now observed may reflect solely that of the donor, which had already destroyed
its Li as in the previous case.
Estimates by \citet{nrbd97} (their Figure~3) show that only a few 
$\times 10^{-2}~M_\sun$ of Li-free material need be accreted to produce a Li-free surface
convective zone. This quantity is compatible with the mass ranges estimated 
above.

\subsection{The non-rotating Li-poor star}

Although this explanation may work for the three line-broadened stars, how should we
interpret the unbroadened one? It is of course possible that this star is rotating
rapidly, but with a low inclination $i$ that renders it unobservable. The same 
explanation could account for the lack of evidence for radial velocity variations,
since the angular momentum acquired during the accretion process would be aligned with
the orbital angular momentum. Detection of a companion to the fourth star would provide support for our
hypothesis.
However, in view of the temperature differences between the hottest three and the 
unbroadened fourth one, we echo the warning of \citet{rbkr01} that more than one
explanation may be required for the formation of the ultra-Li-depleted stars. It is
clearly important to continue the investigation of the remaining sample of 
ultra-Li-depleted objects to determine whether or not they have a common origin.
The chemical signatures of this class of stars, already explored initially
by \citet{nrbd97} and
\citet{rnb98}, will be pursued elsewhere \citep{rgkb02}.

\subsection{Constraints on Li-processing in single stars}

Most single halo
stars have spun down to extremely low velocities over their long histories, so the
existence of high rotation in such old stars requires that a torque has been exerted
on them by some means. 
Mass transfer is one of the most effective ways of achieving this. 
The
compelling evidence for mass transfer having occurred in three of the four
Li-depleted stars in our sample emphasises that 
these stars can tell us nothing about the primordial Li abundance.
If cooler examples of Li-depleted stars 
(distributed over the range of temperatures encompassing the Spite plateau) 
are also found to exhibit
evidence of past mass-transfer episodes that might be responsible for the
destruction of their surface Li, it becomes clear that such objects cannot be
viewed as the high-depletion tail of the same processes that are invoked to
account for the long-term evolution of surface Li abundances in single stars
(e.g., \citet{pwsn01}).
In that case, efforts to study the way single stars
destroy Li as they age will have to avoid
incorporating such objects in their samples.

\section{Summary and concluding remarks} 

The discovery of line broadening, and its likely origin in the
rapid rotation of three of the four ultra-Li-depleted stars, strongly suggests
that these stars have acquired angular momentum from a 
companion star. This possibility is made more plausible by the observation that
all three are single-lined spectroscopic binaries with lower-mass companions.
The angular momentum transfer will have been effected by mass transfer, 
by stable Roche-lobe overflow or 
by a common-envelope phase in the case of the circularized system, 
and wind accretion for the eccentric systems.
Estimates of the specific angular momentum of accreted material indicate
that quite small transfer masses could have been involved, though the 
unknown subsequent spin-down of the accretor prevents us from
assigning definitive values for each star.
In these cases the Li-deficiency could be explained by three 
mechanisms: 
(1) mixing induced by the accretion event;
(2) prior destruction where both stars may have been devoid of Li already 
when mass transfer began;  and
(3) the likelihood that we now observe a surface dominated by accreted material,
which reflects the surface composition of the donor, which had destroyed its own
Li prior to the mass transfer event.

These findings add weight to the
suggestion of \citet{rbkr01} that the ultra-Li-depleted stars, or at least the
hottest subset of them, may be sub-turnoff-mass
counterparts of traditional field blue stragglers. 
Although binarity alone does not provide
evidence of mass transfer, high angular momentum in systems too wide to 
be tidally synchronised does. 

\section{Acknowledgements}

The authors thank the UK Panel for the Allocation of Telescope Time (PATT)
and 
the Director and staff of the Anglo-Australian Telescope (AAT) 
and William Herschel Telescope (WHT) 
for the provision of 
research facilities.
S.G.R. was supported by PPARC grant PPA/O/S/1998/00658.
T.C.B. acknowledges partial support from NSF grants AST 00-98549 and AST
00-58508.
U.C.K. acknowledges partial support from PPARC grant PPA/G/S/1999/00127.
\newpage





NOTE TO EDITOR: PLEASE SET TABLE 1 IN LANDSCAPE MODE OCCUPYING A SINGLE COLUMN,
OR IN PORTRAIT MODE ACROSS TWO-COLUMN WIDTH. THANKS.

\newpage

\begin{deluxetable}{lllllllllllllll}
\tablecaption{Stellar parameters\label{TabA}}
\rotate
\tabletypesize{\tiny}
\tablehead{
\colhead{Object}&\colhead{Inst.}&\colhead{Least-Squares Fit   }&\colhead{ RMS         }&\colhead{$FWHM$$_{\rm +}$}&\colhead{$v$sin$i$ $\pm 3\sigma$   }&\colhead{$T_{\rm eff}$}&\colhead{[Fe/H]}&\colhead{[Fe/H]}&\colhead{$T_{\rm eff}$}&\colhead{[Fe/H]}&\colhead{$P_{\rm orb}$ \tablenotemark{a}}&\colhead{$e$ \tablenotemark{a}}&\colhead{$M_2\sin i$ \tablenotemark{a}} &\colhead{ $i_{\rm NS}$\tablenotemark{b}}\\
      &&\colhead{$FWHM$ (km~s$^{-1}$) =}&\colhead{km~s$^{-1}$}&\colhead{km~s$^{-1}$  }&\colhead{km~s$^{-1}$}&\colhead{phot.   }&\colhead{old  }&\colhead{new  }&\colhead{$\chi$ }&\colhead{$\chi$ }&\colhead{day}&&$M_\sun$&\\
\colhead{(1)   }&\colhead{(2)}&\colhead{(3)                        }&\colhead{(4)          }&\colhead{(5)            }&\colhead{(6)          }&\colhead{(7)     }&\colhead{(8)  }&\colhead{(9)  }&\colhead{(10)}&\colhead{(11)}&\colhead{(12)}&\colhead{(13)}&\colhead{(14)}&\colhead{(15)}\\
}
\startdata
Li-normal&UCLES&$0.0145W_{\rm G}$/m\AA\ + 8.78&0.72\\
CD$-$31$^\circ$19466&
          UCLES&$0.0157W_{\rm G}$/m\AA\ + 8.63&0.61&  0.0& $<$2.2     &5986&--1.89&--1.66&6190&--1.50&\nodata&\nodata&\nodata&\nodata\\
Li-normal&UES  &$0.0126W_{\rm G}$/m\AA\ + 8.55&0.64\\
Wolf 550\tablenotemark{c}
	 &UES  &$0.0106W_{\rm G}$/m\AA\ + 10.62&0.70& 6.3& 5.5$\pm$0.6&6269&--1.66&--1.56&6570&--1.32&688&0.29$\pm$0.04& 0.26 & 10$^{\circ}$\\
BD+51$^\circ$1817&
          UES  &$0.0145W_{\rm G}$/m\AA\ + 12.24&0.78& 9.2& 7.6$\pm$0.3&6345&--1.10&--0.88&6920&--0.48&517&0.043$\pm$0.028& 0.35 & 14$^{\circ}$\\
G202-65  &UES  &$0.0148W_{\rm G}$/m\AA\ + 12.94&0.70&10.1&
8.3$\pm$0.4&6390&--1.50&--1.32&6900&--0.93&167.54&0.15$\pm$0.02& 0.50
& 31$^{\circ}$\\ 
\enddata
\tablenotetext{a}{\citet{cllgm01}}
\tablenotetext{b}{Column (15) gives the inclination $i_{\rm NS}$ {\it if} the 
companion is a neutron star with mass $1.4 M_\sun$.}
\tablenotetext{c}{G66-30}
\end{deluxetable}

\clearpage

\clearpage

\clearpage



\end{document}